\documentclass[conference]{IEEEtran}
\usepackage{cite}

\ifCLASSINFOpdf
\else
\fi
\usepackage[utf8]{inputenc}
\usepackage{booktabs}
\usepackage{amsmath}
\usepackage{amssymb}
\usepackage{paralist}
\usepackage{multirow}
\usepackage{xspace}
\usepackage{color}
\usepackage{xcolor}
\usepackage{ifthen}
\usepackage{url}
\usepackage{fancybox}
\usepackage{enumitem}
\usepackage{listings}
\usepackage{balance}
\usepackage{ifthen}
\usepackage[tight,footnotesize]{subfigure}
\usepackage{graphicx}
\usepackage{amsmath}
\usepackage[linesnumbered]{algorithm2e}
\usepackage{tablefootnote}
\usepackage{comment}
\usepackage{subfig}

\graphicspath{{pics/}}

\definecolor{codegreen}{rgb}{0,0.6,0}
\definecolor{codegray}{rgb}{0.5,0.5,0.5}
\definecolor{codepurple}{rgb}{0.58,0,0.82}
\definecolor{backcolour}{rgb}{0.95,0.95,0.92}

\lstdefinestyle{mystyle}{
	backgroundcolor=\color{backcolour},   
	commentstyle=\color{codegreen},
	keywordstyle=\color{magenta},
	numberstyle=\tiny\color{codegray},
	stringstyle=\color{codepurple},
	basicstyle=\footnotesize,
	breakatwhitespace=false,         
	breaklines=true,                 
	captionpos=b,                    
	keepspaces=true,                 
	numbers=left,                    
	numbersep=2pt,                  
	showspaces=false,                
	showstringspaces=false,
	showtabs=false,                  
	tabsize=2
}

\usepackage[english]{babel}

\DeclareGraphicsExtensions{.pdf,.jpeg,.png}

\begin{document}
%
\title{Extracting Insights from the Topology of the JavaScript Package Ecosystem}



%
\author{\IEEEauthorblockN{Nuttapon Lertwittayatrai\IEEEauthorrefmark{1},
Raula Gaikovina Kula\IEEEauthorrefmark{2},
Saya Onoue\IEEEauthorrefmark{2},
Hideaki Hata\IEEEauthorrefmark{2},\\ 
Arnon Rungsawang\IEEEauthorrefmark{1}, 
Pattara Leelaprute\IEEEauthorrefmark{1} and
Kenichi Matsumoto\IEEEauthorrefmark{2}}
\IEEEauthorblockA{\IEEEauthorrefmark{1}
Faculty of Engineering, Kasetsart University, Bangkok, Thailand\\
\IEEEauthorblockA{\IEEEauthorrefmark{2}
Nara Institute of Science and Technology, Japan\\
Email: nuttapon.l@ku.th, \{arnon.r, pattara.l\}@ku.ac.th} \{raula-k, onoue.saya.og0, hata, matumoto\}@is.naist.jp}
}


\maketitle

\begin{abstract}
Software ecosystems have had a tremendous impact on computing and society, capturing the attention of businesses, researchers, and policy makers alike.
Massive ecosystems like the JavaScript node package manager \texttt{(npm)} is evidence of how packages are readily available for use by software projects.
Due to its high-dimension and complex properties, software ecosystem analysis has been limited.
In this paper, we leverage topological methods in visualize the high-dimensional datasets from a software ecosystem. 
Topological Data Analysis (TDA) is an emerging technique to analyze high-dimensional datasets, which enables us to study the shape of data. 
We generate the npm software ecosystem topology to uncover insights and extract patterns of existing libraries by studying its localities. 
Our real-world example reveals many interesting insights and patterns that describes the shape of a software ecosystem.
\end{abstract}


%
\IEEEpeerreviewmaketitle

\section{Introduction}
Software ecosystems have a tremendous impact on contemporary software development.
Software developers are more likely to rely on third-party libraries from the ecosystem, to gain the benefits of quality, speed to market and ease of use.
One such example of a emerging software ecosystem is the JavaScript Package ecosystem. 
Since inception, the ecosystem has exploded its growth to over half a million\footnote{as of July, 2017 the size of the npm repository reached 475,000 packages.} packages available for its users, making it the biggest and popular Open Source Software ecosystems in recent times.
A ecosystem itself is comprised of many social and technical aspects that can be represented as high-dimensional data.

The datasets gathered from software ecosystems are vastly high-dimensional, noisy and are generally challenging when attempting to identify patterns or insights at a higher level. 
A recent study by Wittern et al. \cite{Wittern2016} investigated some of the dynamics in the study. 
They studied the ecosystem from several perspectives of evolution, popularity and adoption, however, patterns between the different features analyzed separately.
Other related work \cite{Decan2017}, \cite{Kikas2017} focus on the dependencies between the different packages within the npm ecosystem while others have studied from certain social-technical aspects \cite{Constantinou2017}.

In this paper, we apply topological methods to study complex high-dimensional data sets by extracting shapes (patterns) and obtaining insights about them.
Leveraging concepts and algorithms from the mathematics field of Topological Data Analysis (TDA), we provide a geometric representation of complex datasets.
TDA permits the analysis of relationships between related dataset features. 
We illustrate our approach by applying it to a representative sample of packages and six key features that describe the ecosystem topology. 
We are able to extract the following insights from our generated npm ecosystem topology:

\begin{itemize}
\item \textit{The topological shape becomes more refined as more data is added.}
\item \textit{The number of dependencies for a package is a strong feature in the topology.}
\item \textit{Packages that are more likely to be used within ecosystem are located separately from packages meant for application usage outside the ecosystem.}
\item \textit{Top authors of packages tend to release packages intended for usage within the ecosystem itself.}
\item \textit{Packages are not likely to be affiliated to an organizational domain.}
\item \textit{Packages are licensed by a dominant software license (i.e., MIT).}
\item \textit{Popular tagged keywords are common with packages across the topology.}
\end{itemize}
Furthermore, we show how the topology is more insightful than standard alternative methods such as archetypal analysis.
We envision that the study of topology and investigation of additional key features may lead to better understanding of software ecosystems.

\section{Background}
Lum et al. \cite{Lum13} showed how the shape of the topology can be leveraged to extract useful insights.
Lum and colleagues demonstrate how TDA allows exploration of the data, without first having to formulate a query or hypothesis, demonstrating the importance of understanding the “shape” of data in order to extract meaningful insights.

Topology is the field within mathematics that deals with the study of shapes. 
It has its origins in the 18th century, with the work of the Swiss mathematician Leonhard Euler. 
TDA is the result of a concerted effort to adapt topological methods to various applied problems, one of which is the study of large and high dimensional data sets.
In Lum's study, they applied topology to three very different kinds of data, namely gene expression from breast tumors, voting data from the United States House of Representatives and player performance data from the NBA, in each case finding stratifications of the data which are more refined than those that are  produced by standard methods. 

The only other work in which TDA was applied in a software setting was by Costa et al. \cite{Costa2017}.
Using a more complex range of techniques, they concluded that topological analysis might be useful for characterizing software
system behavior early enough and for early characterization of system reliability, that may contribute to software reliability modeling.
In this work, we only focus on the topology mapper algorithm \cite{SPBG:SPBG07:091-100} to generate a topological map of the software ecosystem.



 	\begin{figure}[!t]
		 \center
			 \centering
			\includegraphics[keepaspectratio,scale=0.2]{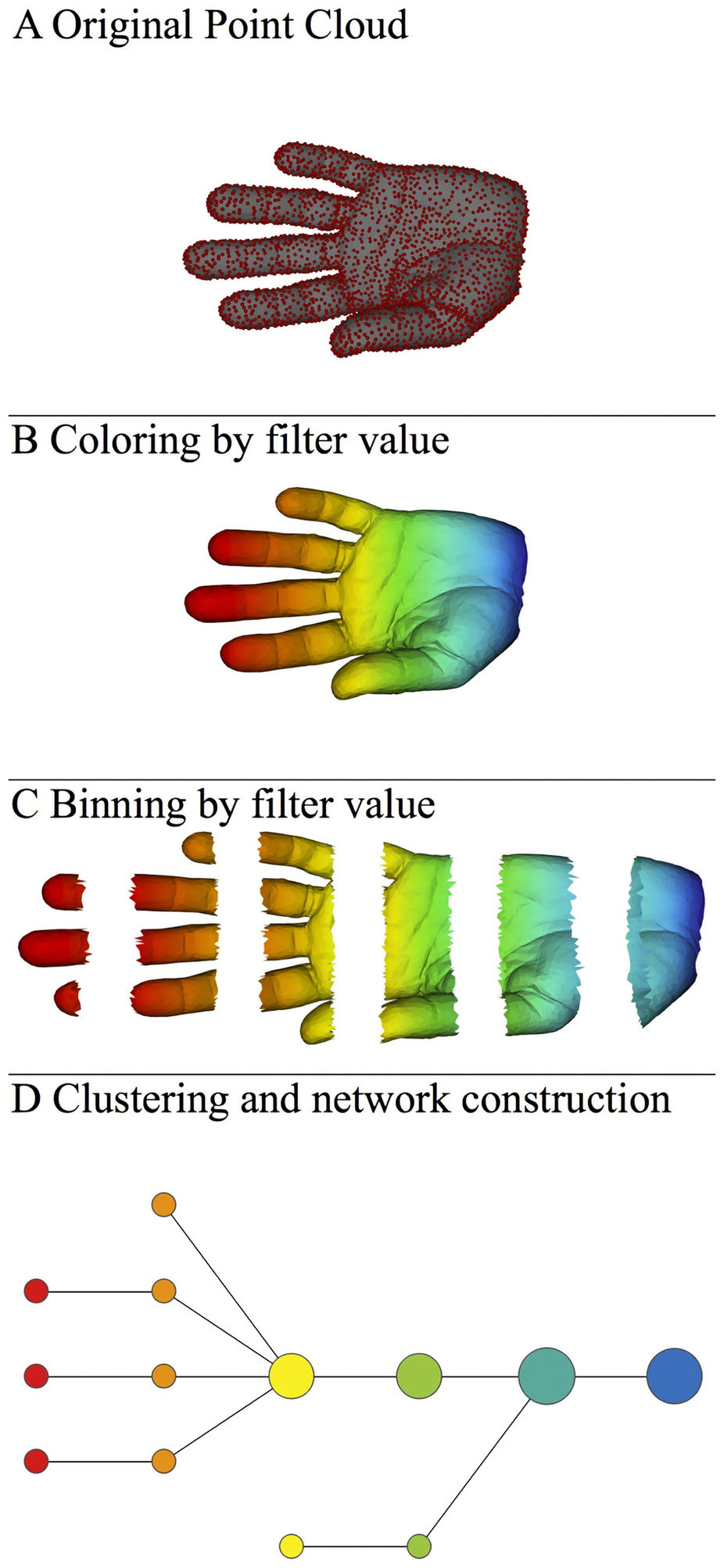}\\
		\caption{Taken from \cite{Lum13}, A 3D object (hand) represented as a point cloud. B) A filter value is applied to the point cloud and the object is now colored by the values of the filter function. C) The data set is binned into overlapping groups. D) Each bin is clustered and a network is built. In this work, each filter is represented by the extracted features of the npm ecosystem.}
		 \label{fig:TDA}
		\end{figure}

\section{A Software Ecosystem Topology}
In this section, we discuss the method by which the topology of an ecosystem is mapped.
We use the definition of software ecosystem as \textit{“a collection of software systems, which are developed and co-evolve in the same environment”} \cite{Lungu2008}. 

\subsection{Topological Mapper Method}

The mapper algorithm \cite{SPBG:SPBG07:091-100} is a method for constructing useful combinatorial
representations of geometric information about high-dimensional point cloud data.
It can be used to reduce high dimensional data sets into some mathematical objects, namely simplicial complexes, with far fewer points that can capture topological and geometric information at a specified resolution.
As shown in Figure \ref{fig:TDA}, instead of acting directly on the data set, it assumes a choice of a filter or combination of filters, which can be viewed as a map to a metric space, and builds an informative representation based on clustering the various subsets of the data set associated the choices of filter values.

\subsection{JavaScript Package Ecosystem Features}
In order to create a point cloud for the ecosystem topology, we first identify six filters (i.e., referred to as features in this paper), which will be indicative of our dataset. 
Mainly based on the work of Wittern \cite{Wittern2016} and other work that studied the software ecosystems \cite{Decan2017}, \cite{Kikas2017}, \cite{Bavota2015}, we identify similar six features that can be used as geometric features of a package.
Similar to Wittern, these six features are present in the meta-file \texttt{package.json}, and are shown in Listing \ref{code:package.json}:

\begin{itemize}
\item \textit{f1- Author}: Name of person who build this package. This indicator should be able to group packages built by the same author. For example from Listing \ref{code:package.json}, in line 11, an author of this package is ``James Halliday''.
\item \textit{f2- Author Domain}: Email domain of person who build this package. This indicator should show packages built by authors from the same organization or company. For example from Listing \ref{code:package.json}, in line 12, an author domain of this package is ``substack.net''.
\item \textit{f3- License}: License tell people know what organization that publish the package how they are permitted to use it. For example from Listing \ref{code:package.json}, in line 5, a license of this package is ``MIT''.
\item \textit{f4- Tagged Keywords}: An array of strings that helps people discover your package as it's listed in \texttt{npm search}. For example from Listing \ref{code:package.json}, in line 16, keywords of this package are ``browswer'', ``requir'', ... ,  ``javascript''.
\item \textit{f5- Version Released}: Version form an identifier that is assumed to be completely unique. Changes to the package should come along with changes to the version. For example from Listing \ref{code:package.json}, in line 3, a version of this package is ``14.4.0''.
\item \textit{f6- Number of Dependencies}: The number of mapped package dependencies to a version range. For example from Listing \ref{code:package.json}, in line 22, dependencies of this package are ``JSONStream'', ``assert'', and ``through''.
\end{itemize}


\begin{lstlisting}[language=xml,
caption={Snippet from the \texttt{package.json} of  the \textit{browserify} package. Some fields are omitted for brevity.},
label=code:package.json]
{
	"name": "browserify",
	"version": "14.4.0",
	"description": "browser-side require() the node way",
	"license": "MIT",
	"repository": {
		"type": "git",
		"url": "http://github.com/substack/node-browserify.git"
	},
	"author": {
		"name": "James Halliday",
		"email": "mail@substack.net",
		"url": "http://substack.net"
	},
	....
	"keywords": [
		"browser",
		"require",
        ....
		"javascript"
	],
	"dependencies": {
		"JSONStream": "^1.0.3",
		"assert": "^1.4.0",
		"through": "^2.3.4"
	},
	....
}\end{lstlisting}

\subsection{Feature Vector Calculation}
One of the complexities of the data is the dimensions within each feature.
To cope with the complexity, we adopt a \textit{Vector Space Model (VSM)} from the Information Retrieval (IR) field to represent the high-dimension of each feature.
For a vector space, we first need a corpus of each of the features.
Suppose we have three packages $P_1$, $P_2$, $P_3$. 
For instance, for the license features \textit{f3}, our corpus will be constructed as follows:
$$
\begin{array}{c|cccc|}
z & P_1 & P_2 & P_3 & ... \\ \hline
MIT & 0 & 1 & 0 & ...  \\
ISC & 1& 0 & 1 & ...  \\
Apache & 0& 0 & 0 & ...  \\
... & ... & ... & ... & ...  \\
\end{array}
$$
In this example, we find that in the license feature, we use the two terms MIT ($z_1$) and ISC ($z_2$).
Thus, we can represent the feature vector for a license as follows:
$$
\vec{P_1}^{\,f3}=
\begin{array}{ccccc}
z_1 & z_2 &  z_3 & ... \\ \hline
0 & 1 & 0 & ... \\  
\end{array}
$$
Note that the corpus 
$m \times n$ matrix whose $i^{th}$, $j^{th}$, is represented by binary 0 or 1 to indicate whether this feature is used by a package or not.
In this example, we can see that the MIT license is used by only $P_2$.
This kind of binary function weighting is used to construct the \textit{$\vec{P_x}^{\,f1}$, $\vec{P_x}^{\,f2}$, $\vec{P_x}^{\,f3}$} vectors.

For the tagged keywords \textit{f4}, we use the \texttt{word2vec} technique as a weighting function instead of the binary function as proposed by Mikolov et al. \cite{DBLP:journals/corr/abs-1301-3781}.
The model is used for learning vector representations of words, such that words that share common contexts in the corpus are located in close proximity to one another (i.e., generated by a similarity score) in that space. 
We use the \texttt{word2vec} function from the gensim python library\footnote{gemsim is a topic modelling library for python. Available at \url{https://radimrehurek.com/gensim/index.html}} to calculate a similarity score between words.
Below, we construct $f4$ for packages $P_1$, $P_2$, $P_3$.


$$
\begin{array}{c|cccc|}
 k & P_1 & P_2 & P_3 & ... \\ \hline
Web & 1 & 0.733261108 & 0 & ...  \\
Http & 0.925269127 & 0.686954796 & 0 & ...  \\%
Console & 0 & 0.889973283 & 0.476446807 & ...  \\%
Server & 0 & 0.76110518 & 1 & ...  \\%
... & ... & ... & ... & ...  \\
\end{array}
$$

In this example, we can visually observer by the similarity scores, that package $P_2$ has a much closer similarity to all of the four keywords \textit{Web} ($k_1$), \textit{Http} ($k_2$), \textit{Console} ($k_3$) and \textit{Server} ($k_4$) than $P_1$ and $P_3$.


The remaining features (i.e., \textit{f5} and \textit{f6}) use a more simplified set of metrics.
For the version released \textit{f5}, we use the release versioning to estimate the current maturity and release of the package. 
In our example `browserify' (i.e., $P_1$) is at current version 14.4.0. 
Therefore, we represent this package as ${P_1}^{f5}=14.4$.
Similarly, we use a count of the dependencies as a measure of how dependent a package is on the ecosystem. 
In this example, we find that browserify lists 3 dependencies (JSONStream, assert and through), hence 
${P_1}^{f6}=3$.
Finally, we combine all the feature vectors to end with a single vector for each package. 
For instance:
\begin{equation*}
\vec{P_x} = \vec{P_x}^{\,f1} \wedge \vec{P_x}^{\,f2} \wedge \vec{P_x}^{\,f3} \wedge \vec{P_x}^{\,f4} \wedge {P_x}^{f5} \wedge {P_x}^{f6}
\end{equation*}

Since each vector is a matrix, it is important to note that size of the dimension for each feature is dependent on the size of the terms ($z_1, ..., z_x$) in each feature. 
The key advantage of our TDA approach is the ability to process and visualize these types of high-dimensional datasets.

\subsection{Data Collection and Topology Representation}

\begin{table}[t]
	\centering
	\caption{Summary of Data Collected}
    \label{tab:data_collected}
	\begin{tabular}{l|cc|c|c}
		\hline\hline
		& Dataset Statistics  \\ \hline
        Snapshot Date & July-1st-2016 $\sim$ July-15th-2016\\
        \# Collected Packages (after filtering) & 72,650 \\
        \# sample packages (generate topology) & 10,000 \\
		\hline
	\end{tabular}
\end{table}

To evaluate our topology methodology, we used a sample of the npm ecosystem.
In this section, we will describe the data collection and visualization analysis.

\subsubsection{Data Extraction and Preprocessing}

As shown in Listing \ref{code:package.json}, we are able to extract all our metrics by mining the \texttt{package.json} from each package.
Using the same method from prior work, we randomly selected and mined 151,100 JavaScript npm packages.
To improve our data collection, we only select packages that include all the features needed. 

To extract each dimension, we used python scripts with the following libraries.
For the \texttt{word2vec} analysis, we used a standard threshold (i.e., 500 words) as a base for the  algorithm.

\begin{table}[t]
	\centering
	\caption{Size of dimension for each granularity of $\vec{P_x}$. Note the Top features are related to $f1$,$f2$,$f3$ and $f4$ features.}
    \label{tab:scalability}
	\begin{tabular}{l|cccc}
		\hline\hline
		Degree & \# Dimensions & Data Size & Topology Generation Time \\ \hline
		Top 20 &  82 x 10,000 & 1.9 MB & 10.53 minutes \\
		Top 50 & 202 x 10,000 & 4.7 MB & 18.47 minutes \\
		Top 100 & 402 x 10,000 & 9.4 MB & 39.42 minutes \\
		Top 1000 & 3,380 x 10,000 & 86.8 MB & 52.16 minutes \\
		\hline
	\end{tabular}
\end{table}



\begin{figure*}[t]
\centering
	\begin{tabular}{cccc}
			\hspace{-10mm}
			\begin{minipage}{0.25\linewidth}
			\centering
            \includegraphics[keepaspectratio,scale=0.15,angle=0]{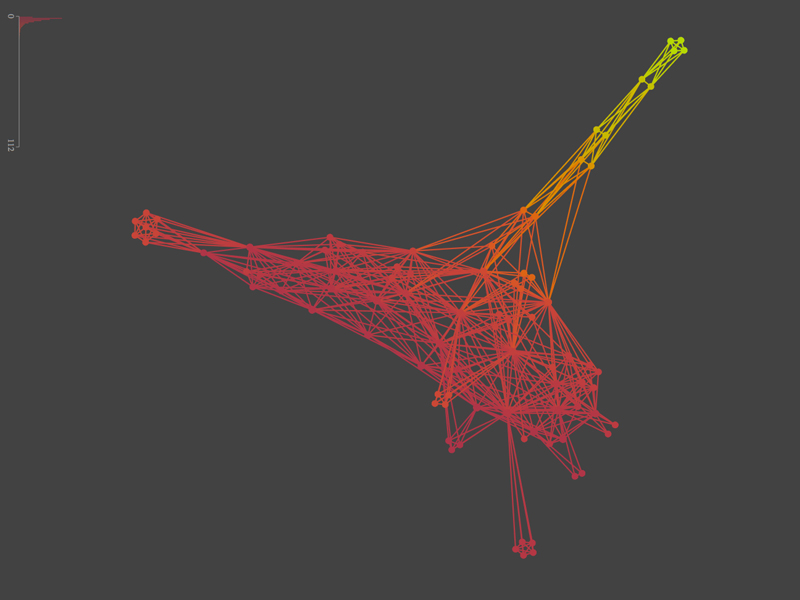}
			\\
            (a) Top 20
			\end{minipage}
            &
			\begin{minipage}{0.25\linewidth}
			\centering
            	\includegraphics[keepaspectratio,scale=0.15,angle=0]{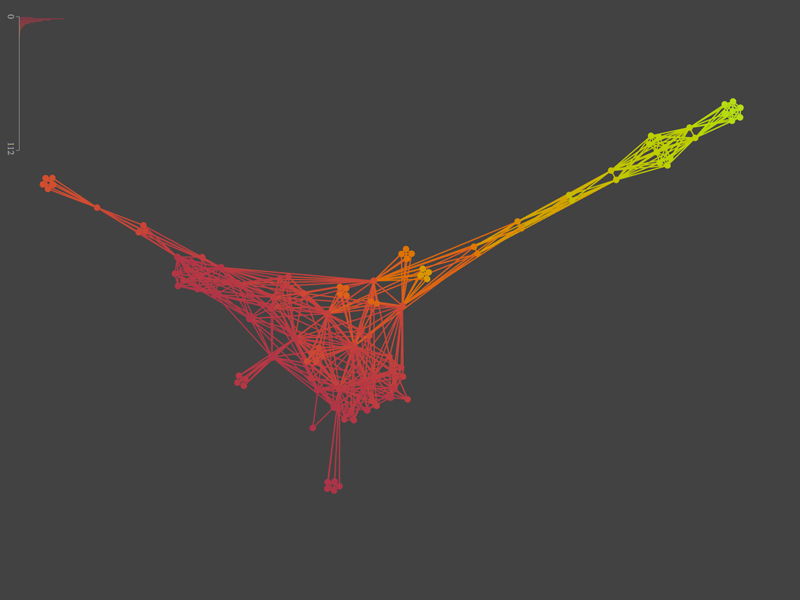}
            \\
            (b) Top 50
			\end{minipage}
           	&
			\begin{minipage}{0.25\linewidth}
			\centering
            	\includegraphics[keepaspectratio,scale=0.17,angle=0]{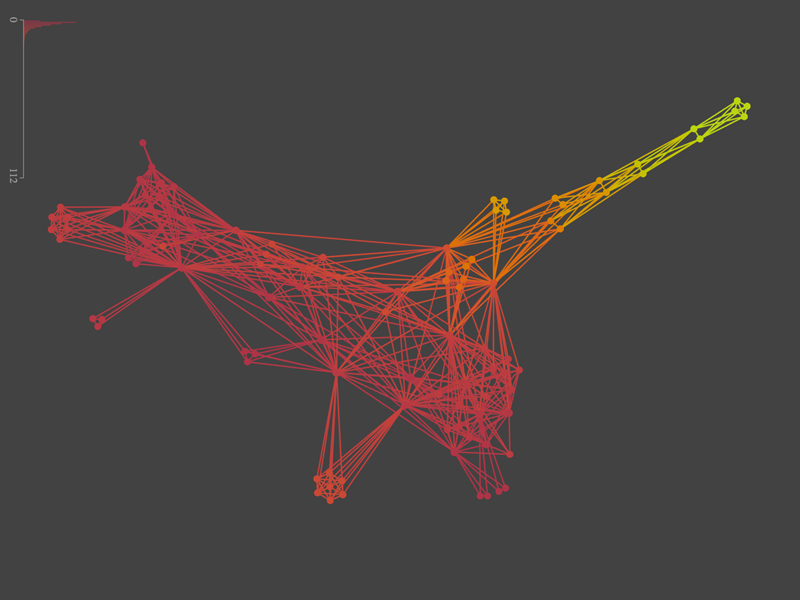}
            \\
            (c) Top 100
			\end{minipage}
			&
			\begin{minipage}{0.25\linewidth}
			\centering
            	\includegraphics[keepaspectratio,scale=0.15,angle=0]{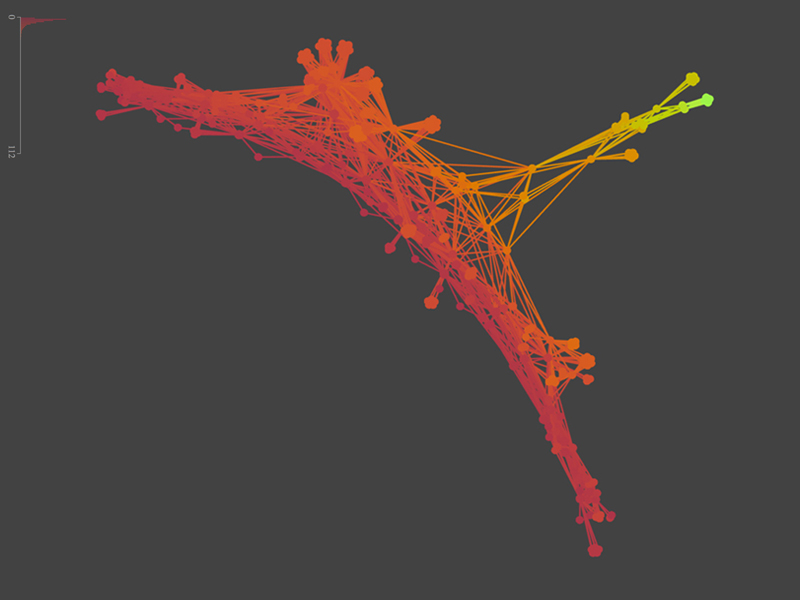}
            \\
            (d) Top 1000
			\end{minipage}
			\end{tabular}
			\\
		\caption{Summary view of the JavaScript Package topology at different granularities. We find that the shape evolves, yet is able to maintain its key points.}
		 \label{fig:evolution}
	\end{figure*}

\subsubsection{Using the Mapper algorithm in TDA}
We use the Knotter tool\footnote{\url{https://github.com/rosinality/knotter}}, which is an implementation of mapper algorithm for TDA \cite{SPBG:SPBG07:091-100}.
The method provides a common framework which includes the notions of density clustering trees, disconnectivity graphs, and Reeb graphs, but which substantially generalizes all three. 
We use the t-Distributed Stochastic Neighbor Embedding (t-SNE) \cite{Maaten2008}, a technique for dimensionality reduction and clustering, and our defined features as the filters for the visualization construction.


We use different layers of granularity of the features.
Table \ref{tab:scalability} shows the scale of the high-dimensional features (i.e., $f1, f2, f3, f4$) at the three levels of Top 20, Top 50, Top 100 and Top 1,000. 
This intention is to understand whether the key features can be seen at the high-dimensions of the data analysis.
Due to the limitations of the tool, we were only able to load up to 10,000 packages (i.e., with loading times of over 50 minutes) into the topology.
Table \ref{tab:scalability} details the data size and topology generation loading times. 

 	\begin{figure*}[!t]
		 \center
			 \centering
			\includegraphics[keepaspectratio,scale=0.5]{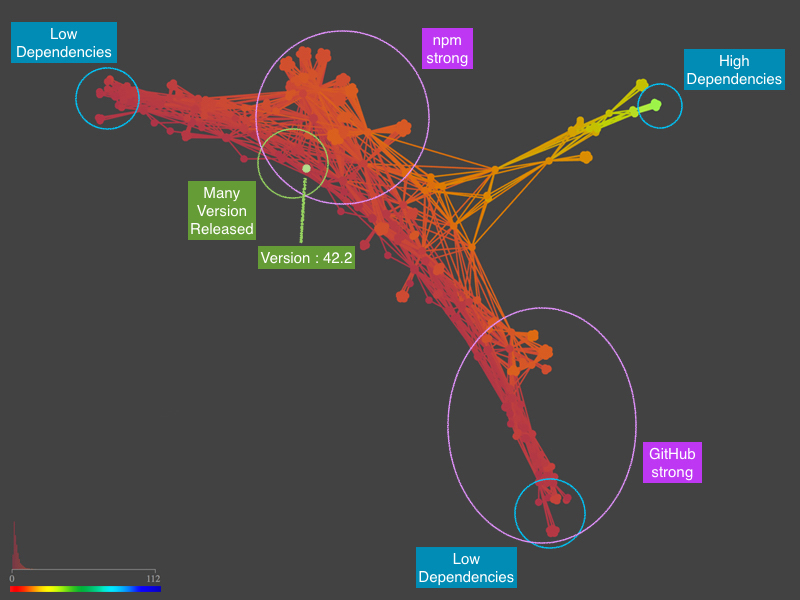}\\
		\caption{The npm Ecosystem Topology. Color is related to the \textit{f6} feature.}
		 \label{fig:overview}
		\end{figure*}

\subsection{Analysis and Interpretation of the Topology}

Our analysis is by a visual analysis and identification of important features on the shape of the topology.
We interpret the results of the topology using two levels of analysis: 

\subsubsection{Topology Overview and Shape Analysis}
We analyze the topology of the ecosystem, analyzing the shape of the data over the different granularities (i.e., Top 20, Top 50, Top 100, Top 1,000).
Furthermore, we then investigate the dominant features that outline the shape of the topology.

\begin{table}[t]
	\centering
	\caption{`GitHub strong' vs. `Npm strong' tagged keywords ($f3$) as discussed by Wittern et. al \cite{Wittern2016}.}
    \label{tab:strong}
	\begin{tabular}{c|cc}
		\hline\hline
		`GitHub strong' & `npm strong' \\ \hline
		gruntplugin & util & \\
		gulpplugin & array & \\
		express & buffer & \\
		react & string & \\
		authenticate & file & \\   
        \hline
	\end{tabular}
\end{table}

One particular analysis is the categorization of tagged keywords. 
Shown in Table \ref{tab:strong}, Wittern et al. \cite{Wittern2016} found a set of keywords that were likely to be related to either applications (i.e., \texttt{GitHub strong}) or the npm package ecosystem (i.e., \texttt{npm strong}). 
As part of our analysis, we would like to identify the locations of libraries that use these sets of keywords.

\subsubsection{ Deeper Topology Feature Analysis}
We analyze each feature to identify some interesting observations and their relationships to the other features.
Our method is to identify the locations of the most frequent occurring terms (i.e., top 5 $z_x$) of each feature.  
For instance, in reference to the authors ($f1$), we will map the libraries that belong to top 5 authors of npm packages.
In the case study, we specifically look at the Author ($f1$), Author Domain ($f2$), License ($f3$) and Tagged Keywords ($f4$) features of the topology.

\section{Results}
In this section, we discuss our results in terms of (1) the topology overview and (2) topology features for our constructed npm ecosystem topology.

\subsection{JavaScript Package Ecosystem Topology}
\vspace{2mm}
\begin{quote}
\textit{``The topological shape becomes more refined as more data is added''}
\end{quote}

Figure \ref{fig:evolution} depicts the shape of the data at the different granuality of dimension levels (i.e., Top 20, Top 50, Top 100 and Top 1,000).
We can see for the figures that each shape is geometrically different, however, the key points of the shape are still the same. 
This result indicates that libraries with these high features are apparent in the Top 20. 
However, an argument could be said that the shape is more refined as more data is added.
By refined, we mean that the extremely points of the data (i.e., represented by the edges of the shape) become more apparent.
It is because of this reason, that we decided to perform the rest of our analysis at the Top 1,000.

Figure \ref{fig:overview} depicts a detailed analysis of important points in the topology mapped to some of the feature attributes. 
From this figure, we are able to extract the following insights.

\vspace{2mm}
\begin{quote}
\textit{``The number of package dependencies is a strong feature in the topology''}
\end{quote}

We find that the shape of the topology is influenced by the number of dependencies adopted by a package ($f6$).
This is clearly highlighted by the top-right in the shape. 
Actually, such high dependency packages may risk becoming blacklisted\footnote{a blog on these types of packages and their impact to the ecosystem is at \url{https://github.com/jfhbrook/hoarders/issues/2}} due to the debate of whether or not it is simply hoarding other packages.
Conversely, the other two points show a lower set of dependencies.

\begin{table}[!t]
	\centering
	\caption{Top 5 ranking (highest frequency) for each Feature}
    \label{tab:f56}
	\begin{tabular}{c|ccc}
		\hline\hline
		Frequency & Versions (f5) & \# Dependencies (f6) \\ \hline
		1 & 0.0 (2,214) & cordova-plugin-require-bluetoothle (112) \\
		2 & 1.0 (2,202) & npm (85)\\
		3 & 0.1 (1,682) & gtb (62) \\
		4 & 0.2 (675) & mikser (61) \\
		5 & 1.1 (579) & react-setup (61)\\ 
        ... & 42.2 (1) & ...\\ 
        \hline
	\end{tabular}
\end{table}

\vspace{2mm}
\begin{quote}
\textit{``Packages that are more likely to be used within ecosystem are located separately from packages meant for application usage outside the ecosystem''}
\end{quote}

Figure \ref{fig:overview} clearly shows that packages containing the \texttt{`npm strong'} (i.e., ecosystem-use) libraries are located apart from libraries that are \texttt{`GitHub strong'} (i.e., application-use).
We found that the released versions was not an important feature in the topology. 
However, as shown in Figure \ref{fig:overview}, we can identify that package that had the most releases is located near the \texttt{`npm strong'} libraries.
In fact, we found this package to be \textit{ydr-utils}, which is indeed used specialized packages within the npm ecosystem\footnote{inspection of the readme.md file shows that it is used by a specialized set of npm packages \url{https://github.com/cloudcome/nodejs-ydr-utils}}.

\subsection{Topology Features}
\begin{table*}[!t]
	\centering
	\caption{Top 5 ranking (most frequent terms ($z_x$)) for each Feature}
    \label{tab:f1234}
	\begin{tabular}{c|ccccccc}
		\hline\hline
		Frequency Rank & Author (f1) & Author Domain (f2) & License (f3) & Tagged Keywords (f4) \\ \hline
		1 & Author 1 (437) & gmail.com (7,576) & MIT (6,715) & react (3,084) && \\
		2 & Author 2 (436) & substack.net (328) & ISC (1,191) & api (2,984) &&\\
		3 & Author 3 (328) & outlook.com (173) & APACHE-2.0 (950) & yeoman-generator (2280) &&\\
		4 & Author 4 (275) & gmx.de (134) & BSD-2-CLAUSE (524) & cli (2,210) &&\\
		5 & Author 5 (265) & qq.com (97) & BSD-3-CLAUSE (452) & css (2,173) &&\\
		\hline
	\end{tabular}
\end{table*}

\begin{figure*}[p]
\centering
\hfil
\hspace*{-6mm}
 \begin{tabular}{cc}
	\begin{minipage}{0.5\linewidth}
	\centering
    \includegraphics[keepaspectratio,scale=0.4,angle=90]{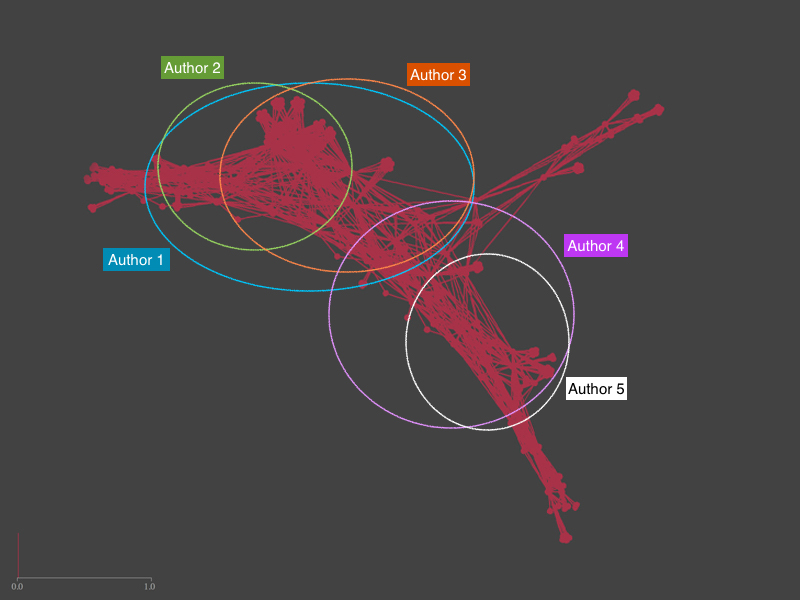}
	\\
    \caption{Author Top 1000}
	\label{fig:f1}
 	\end{minipage}
	&
 	\begin{minipage}{0.5\linewidth}
	\centering
    \includegraphics[keepaspectratio,scale=0.4,angle=90]{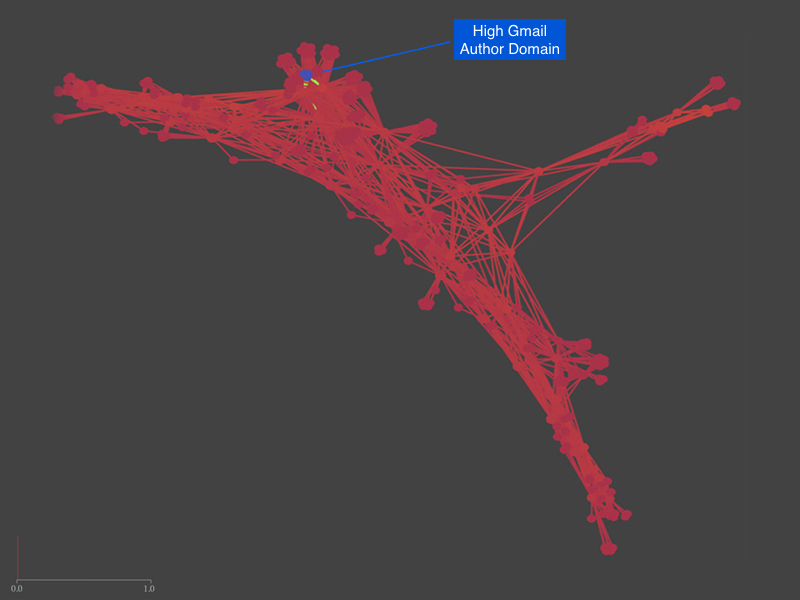}
	\\
    \caption{Author Domain Top 1000}
	\label{fig:f2}
	\end{minipage}
\end{tabular}
\hspace*{-6mm}
 \begin{tabular}{cc}
	\begin{minipage}{0.5\linewidth}
	\centering
    \includegraphics[keepaspectratio,scale=0.4,angle=90]{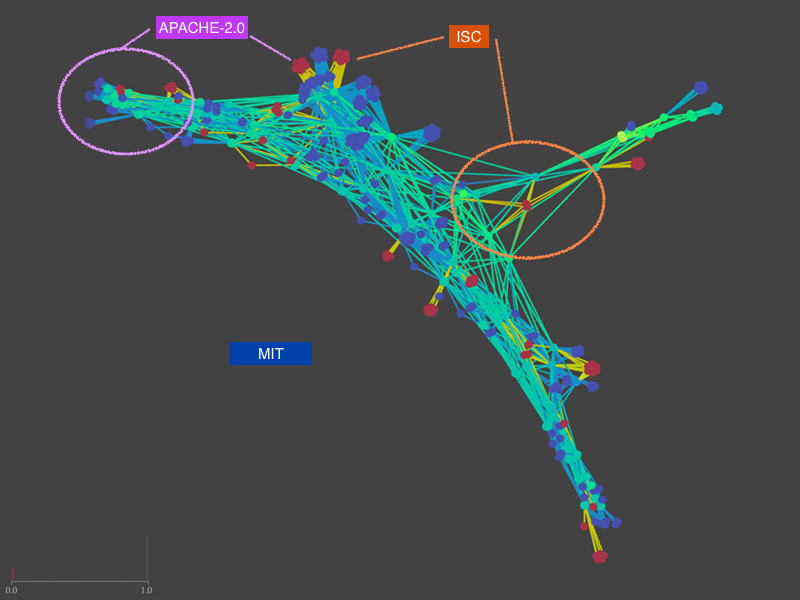}
	\\
    \caption{License Top 1000}
	\label{fig:f3}
	\end{minipage}
	&
    \begin{minipage}{0.5\linewidth}
	\centering
	\includegraphics[keepaspectratio,scale=0.4,angle=90]{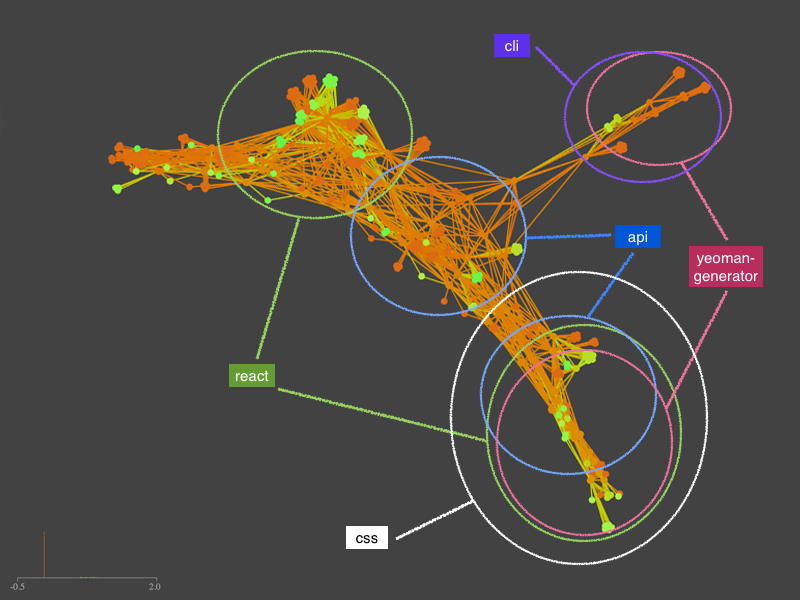}
	\\
    \caption{Tagged Keywords Top 1000}
	\label{fig:f4}
	\end{minipage}
\end{tabular}
\end{figure*}
 	\begin{figure*}[!t]
		 \center
		\hspace{-5mm}
			 \centering	\includegraphics[keepaspectratio,scale=0.545]{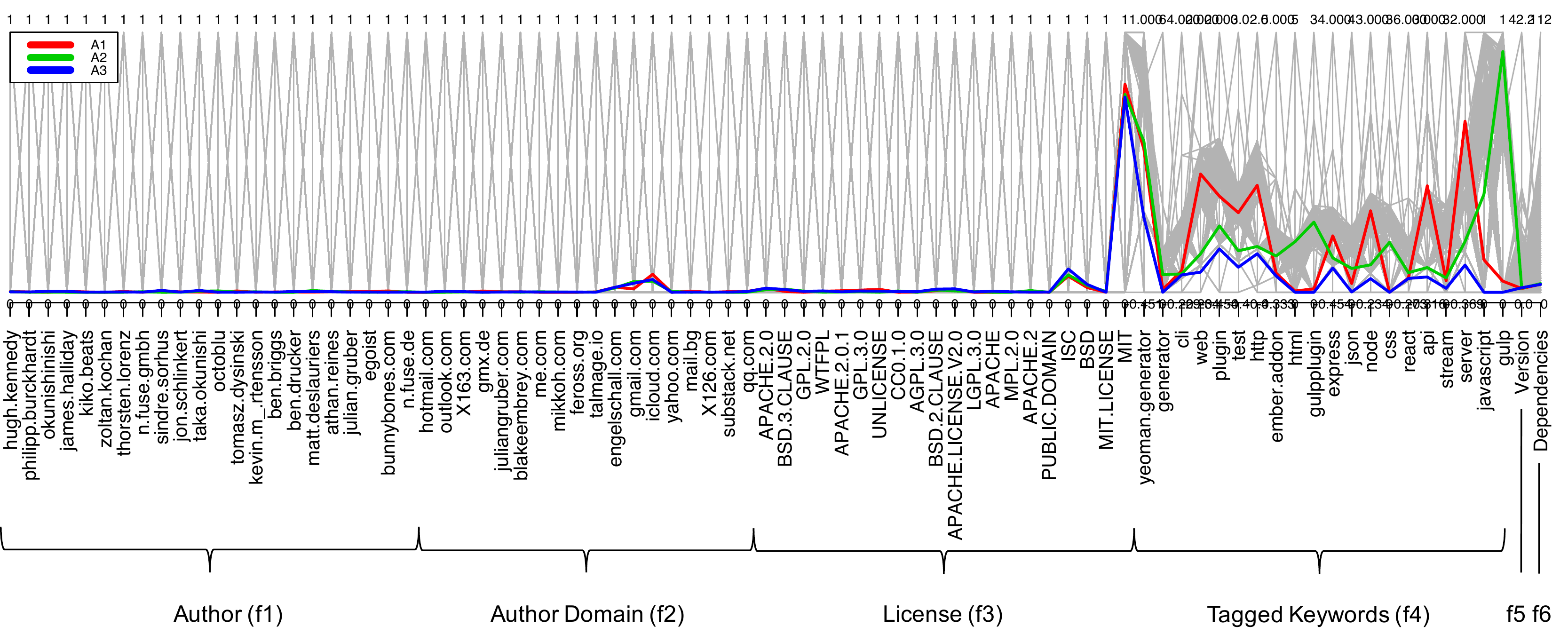}\\
		\caption{The figure shows parallel coordinate plot for the Top 20 npm dataset. The red line is archetype 1 (A1), the green line is archetype 2 (A2) and the blue line is archetype 3 (A3).}
		 \label{fig:pcplot}
		\end{figure*}
\begin{table*}[t]
	\centering
	\caption{Packages that were identified as close to the extreme points of each archetype A1, A2 and A3.}
    \label{tab:aa}
	\begin{tabular}{c|ll}
		\hline\hline
		Archetype & Identified Packages\\ \hline
		A1 & tar-parse, turtle-run, marked-sanitized, haversort, bmxplayjs & \\
        A2 & statsd-influxdb-backend, ardeidae, demo-blog-system, git-ssb-web, social-media-resolver & \\
        A3 & stream-viz, programify, polyclay-couch, meshblu-core-task-check-update-device-is-valid, apidoc-almond & \\
        \hline
	\end{tabular}
\end{table*}
 	\begin{figure}[!t]
		 \center
			 \centering
			\includegraphics[keepaspectratio,scale=0.4]{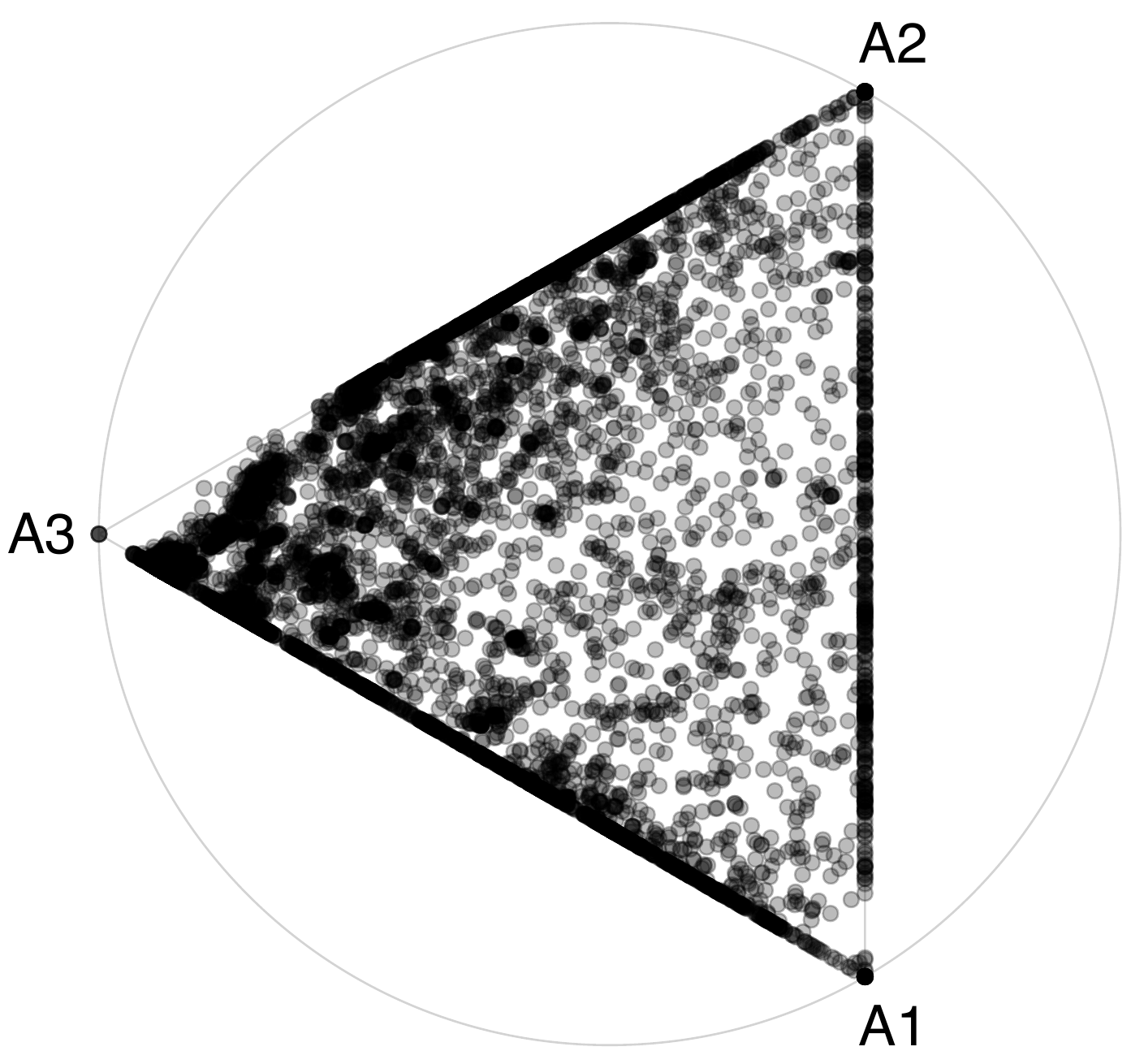}\\
		\caption{A simplexplot that shows the a triangle plot that represents a package in the ecosystem. Note the extreme points represent each archetype.}
		 \label{fig:simplexplot}
		\end{figure}
Figures \ref{fig:f1}, \ref{fig:f2}, \ref{fig:f3}, \ref{fig:f4} shows detailed feature information related to authors ($f1$), author domains ($f2$), license ($f3$) and tagged keywords ($f4$).   
Tables \ref{tab:f1234} and \ref{tab:f56} supplement these Figures by showing the Top 5 most frequent terms for all features.
Drawing from all the presented information, we are able to make the following observations:
\vspace{2mm}
\begin{quote}
\textit{``Top authors of packages tend to release packages intended for usage within the ecosystem itself''}
\end{quote}
Figure \ref{fig:f1} and Table \ref{tab:f1234} shows that the Top 3 authors (i.e., authors 1 with 437 packages, author 2 with 436 packages and author 3 with 328 packages) are located in the same location as the `npm strong' libraries, thus providing evidence that the Top 3 authors were more likely to develop packages for the npm ecosystem.
However, the Top $4^{th}$ (i.e., with 275 packages) and $5^{th}$ (with 265 packages) authors develop packages aimed for applications (i.e., `GitHub strong').

\vspace{2mm}
\begin{quote}
\textit{``Packages are not likely to be affiliated to an organizational domain''}
\end{quote}

Figure \ref{fig:f2} and Table \ref{tab:f1234} shows that the \texttt{gmail} domain is significantly (i.e., showing 7,576 packages) used by authors of npm packages.
The second most used domain is for the \texttt{substack.com} domain\footnote{author GitHub profile at \url{https://github.com/substack}}, which belongs to the Author 3.
This evidence suggests that many of the packages are indeed contributed by individuals and not by developers that are represented by a single organization.
Furthermore, packages created by the authors with no affiliation are more likely to contribute packages that will be used by the npm ecosystem.

\vspace{2mm}
\begin{quote}
\textit{``Packages are licensed by a dominant software license (i.e., MIT)''}
\end{quote}

Figure \ref{fig:f3} and Table \ref{tab:f1234} show the MIT license to be the most widespread license (with 6,715) used for packages in the npm ecosystem.
The next closest popular licenses include the ISC (i.e., with 1,191) and APACHE-2.0 (i.e., with 950).

\vspace{2mm}
\begin{quote}
\textit{``Popular tagged keywords are common with packages across the topology''}
\end{quote}

Figure \ref{fig:f4} and Table \ref{tab:f1234} illustrates how the most frequent keywords (i.e., react, api, yeoman-generator, cli and css) are used across the topology of packages.
This result provides evidence that the individual tagged keywords are generic, therefore not strong indicators for a software ecosystem topology.


\section{Comparison with Archetypal Analysis}




An archetypal analysis is a statistical method that synthesizes a set of multivariate observations through a few \textit{\textbf{archetypes}}, which lie on the boundary of the data scatter and represent pure individual types \cite{Porzio:2008:UAB:1416593.1416598}.
We use the archetypal analysis to compare how useful the topology is for visualizing and analyzing high-dimensional data.
In detail, archetypal analysis describes individual data points based on the distance from extreme points, archetypes.
We used \textsf{R} package \textsf{archetypes} \cite{Eugster:Leisch:2009:JSSOBK:v30i08} for the analysis of the Top 20 npm dataset (See Figure \ref{fig:evolution} and Table \ref{tab:scalability}) used in the previous TDA analysis.
From the ``elbow criterion'' with the curve of the residual sum of squares (RSS), $k = 3$ is determined as the number of archetypes.

Figure \ref{fig:pcplot} presents a parallel coordinate plot of all packages and each line represents one package with all feature values.
The three colored lines are archetypes in this data (red is Archetype 1, green is Archetype 2, and blue is Archetype 3).
We find that the keywords are the strongest feature indicators.
In addition, Table \ref{tab:aa} presents some of the actual packages close to obtained three archetypes.
From our analysis, we were able to qualitatively summarize our findings related to tagged keywords ($f4$):
\vspace{2mm}
\begin{itemize}
\item \textit{Archetype 1} (A1) has packages that contains keywords, such as \textit{web, plugin, test, http, express, node, api and server.}
\item \textit{Archetype 2} (A2) has packages that contains keywords like\textit{ html, gulpplugin, css, javascript and gulp. }
\item \textit{Archetype 3} (A3) has lower packages compared to the other two archetypes.
\end{itemize}

Figure \ref{fig:simplexplot} shows triangular graph which is plotted result of archetype analysis, we a dot representing a package in the npm ecosystem.
From the figure, we can observe that the packages are widely distributed among three archetypes.
This provides us evidence to argue that the topology provides more insights and patterns as compared to this archetypal analysis.
One of the explanations is due to the limited number of archetypes. 
We observe that summarizing these small number of representations is prone to loose
some information, making the raw data too complex for interpretation. 
With respect to the topology, investigation of the topology shape provides use with a more flexible analysis while reducing the dataset. 

\section{Discussion}

In this section, we discuss the implications and then follow up with the threats to our study. 
This includes the possible applications of how developers can leverage the software ecosystem topology.

\subsection{Implications}
\label{sec:imp}
We discuss two benefits where understanding the topology is beneficial for both practitioners and researchers alike.
The first is for searching and selection of components (i.e., packages) within the ecosystem. 
This has been some work that empirically the update and dependency relationships within the ecosystem \cite{Decan2017,Kikas2017,KulaEMSE2017,Ihara2017,Ishio2017}.
We envision that based on a query of features, a developer should be able assess their options from the topology and make a more informed decision on similar or recommended libraries. 
The topology may also be used as a guide for novice developers.
For example, a developer can use the topology as a guide to some of the more common practices (i.e., licensing the package under MIT). 
For future work, we would like to further explore how an ecosystem topology can be leveraged to search and recommend similar or useful packages for a developer.

The second benefit is related to the sustainability and scalability of a software ecosystem.
We believe that such methods such as ecosystem topology provides us a more empirical means to assess various inconspicuous patterns within an ecosystem. 
For instance, the topology can reveal the location of packages that are either made for the npm ecosystem or for application usage. 
Such patterns may become indicators of the ecosystem health.
For future work, we would like to explore additional features, especially the more social (i.e., contributors and open source development organization activities ) or technical aspects (i.e., source code evolution) of packages within the ecosystem.
Furthermore, we would like to study the evolution of the ecosystem topology as an additional future work. 

There are many challenges related to software ecosystems.
Work by Serebrenik and Mens \cite{Serebrenik2015} grouped these challenges as: \textit{ Architecture and Design, Governance, Dynamics and Evolution, Data Analytics, Domain-Specific Ecosystems Solutions, and Ecosystems Analysis.}
We believe that topology analysis may prove to be useful in addressing some of these issues at the higher level.
Other future avenues for research are related to the extension of our method and techniques.
We find that the topology provides a holistic method to visualize and compare these features at a higher-level. 
In this work we only implement the topology method (i.e., TDA mapper) within the TDA field. 
Future work may include a more in-depth analysis using other TDA concepts such as persistence homology.

Our overall vision is towards a more concrete means of automated library recommendations and categorizations.
We believe that this study a step in this direction. 
Hence, future work will include a more deeper look at the TDA and provide more realistic and useful library categorized that are actionable for software developers.

\subsection{Threats to Validity}

\subsubsection{Construct}
This validity is concerned with threats to the construction of the topology, which is the selection of the features.
We understand that there is a plethora of other much more stronger indicators that could be used in the topology.
In this work, we use the six features that are popularly used in prior works \cite{Wittern2016}, \cite{Decan2017}, \cite{Kikas2017}, \cite{Bavota2015}.
As discussed in the prior section (Section \ref{sec:imp}), we plan to expand our feature list in the future.

\subsubsection{Internal}
This validity is related to the accuracy of the data collected and tools used in the experiments.
For the topology generation, we randomly selected 10,000 npm projects (discarding any package that was missing any of the features) to generate our topology.
We understand that with the rate by which an ecosystem changes, that the results may quickly become outdated. 
However, based on the different granularity (Top 20, Top 50, Top 100, Top 1,000) we believe that the shape of the data may change but the structure may still resemble our current topology.
To validate this, we would have to experiment with much more data.

A minor threat to our study is our \textit{vsm} formulation for each package.
For instance, we use the \texttt{word2vec} technique for the tagged keyword feature. 
Our main rational is that the word2vec provides a more robust techniques compared to the basic binary technique. 
Although this is not empirically evaluated, we are confident that the results are representative.
The second threat to validity is the accuracy of the tool used to generate the topology.
Our initial experiments included other topological mapper tools such as the Kepler Mapper\footnote{\url{https://github.com/MLWave/kepler-mapper}} and a \textsf{R} TDA package\footnote{\url{https://cran.r-project.org/web/packages/TDA/index.html}}.
However, we find that the knotter tool is the more versatile tool with multi-function and is able to analyze a large-scale data.

\subsubsection{External}
The external validity refers to the generalization of our results. 
Currently, we agree that the results may only be specific to the npm ecosystem. Therefore, as future work, we would like to explore other software ecosystems, thus comparing the relative similarities and differences between the ecosystem topology shape. 
We are, however, are confident that our sample data is representative of the current ecosystem topology of npm packages.

\section{Related Work}
In this section, we briefly present literature related to analysis of (1) software ecosystems in terms of package dependencies and (2) application of TDA to other domains. 

\subsection{Software Ecosystems}
In literature, a software ecosystem can be defined from a technical \cite{Lungu2008}, business \cite{Jansen2013} viewpoints. 
Our work is mainly from a technical viewpoint, with our features extracted from the meta files of the package.json files.
From a holistic viewpoint work by German et al. \cite{German2013} and Wittern et al. \cite{Wittern2016} empirically studied the different dynamics for the \textsf{R} and JavaScript npm package ecosystems.

There has been many works that have studied relationships among components within the ecosystem. 
The most common is the evolution and update of dependency relationships within the ecosystem \cite{Decan2017,Kikas2017,Bavota2015,KulaEMSE2017,German2013}, and the npm and other software ecosystems such as R, Maven and Ruby \cite{Raemaekers2012}.
Most of these works focus on dependency relationships between the software packages and their evolution.
In particular, work by Bavota \cite{Bavota2015} used more features in their analysis.
In this work, we combine all features to understand at a higher level how packages in the ecosystem are related to each other. 

There has been prior work on visualizations have attempted to show relationships within the ecosystem \cite{KulaVISSOFT2014,Yano2015}, but have only focused on the dependency feature of the ecosystem. 
For future work, we would like to expand these features to cover the more social aspects of \cite{Constantinou2017}, \cite{Onoue2014,Onoue2016} of the project contributors. 
Likewise, source code metrics such as the code complexity and number of functions could be also added to the topology.


\subsection{Application of TDA topology in other fields}
In addition to the insights from the study of Lum et al.\cite{Lum13}, 
TDA is extensively used in the medical field.
The topology has been used to identify patterns in medical information.
For example, Nicolau et al. used TDA to identify a subgroup of breast cancers with a unique mutational profile \cite{Nicolau2011}.
Nielson et al. used TDA to discover pre-clinical spinal cord injury and traumatic brain injuries \cite{Nielson2015}.
TDA has also been used to study infectious disease, \cite{Louie2016}, the Escherichia coli O157:H7 \cite{Ibekwe2014} and type 2 diabetes \cite{Li2015}.
We believe our work will be the start of such similar types of analysis to understand different aspects of software ecosystems.

\section{Conclusion}
In this paper, we present a new approach to analyze and explore high-dimensional and complexity of software ecosystem using the topological data analysis approach.
Applied to real-world high-dimensional complex dataset of the JavaScript Package ecosystem using six features of the ecosystem. 
Our results show that the a software topology is possible and useful to understand higher level relationships within the ecosystem.
The results also show that topology analysis is more insightful that alternative traditional statistical methods, especially with complex data.

For future work, we plan to expand the TDA topology features and techniques.
In addition, we would like to explore the shape of other software ecosystems.
We envision that the study of topology and investigation of is beneficial for both practitioners and researchers, providing a better understanding of software ecosystems.


\section*{Acknowledgment}
This work has been supported by JSPS KAKENHI Grant Number 16H05857.




%

\balance


\begin{thebibliography}{10}
\providecommand{\url}[1]{#1}
\csname url@samestyle\endcsname
\providecommand{\newblock}{\relax}
\providecommand{\bibinfo}[2]{#2}
\providecommand{\BIBentrySTDinterwordspacing}{\spaceskip=0pt\relax}
\providecommand{\BIBentryALTinterwordstretchfactor}{4}
\providecommand{\BIBentryALTinterwordspacing}{\spaceskip=\fontdimen2\font plus
\BIBentryALTinterwordstretchfactor\fontdimen3\font minus
  \fontdimen4\font\relax}
\providecommand{\BIBforeignlanguage}[2]{{%
\expandafter\ifx\csname l@#1\endcsname\relax
\typeout{** WARNING: IEEEtran.bst: No hyphenation pattern has been}%
\typeout{** loaded for the language `#1'. Using the pattern for}%
\typeout{** the default language instead.}%
\else
\language=\csname l@#1\endcsname
\fi
#2}}
\providecommand{\BIBdecl}{\relax}
\BIBdecl

\bibitem{Wittern2016}
E.~Wittern, P.~Suter, and S.~Rajagopalan, ``{A look at the dynamics of the
  JavaScript package ecosystem},'' in \emph{Proceedings of the 13th
  International Workshop on Mining Software Repositories - MSR '16}.\hskip 1em
  plus 0.5em minus 0.4em\relax New York, New York, USA: ACM Press, 2016, pp.
  351--361.

\bibitem{Decan2017}
A.~Decan, T.~Mens, and M.~Claes, ``{An empirical comparison of dependency
  issues in OSS packaging ecosystems},'' in \emph{2017 IEEE 24th International
  Conference on Software Analysis, Evolution and Reengineering (SANER)}.\hskip
  1em plus 0.5em minus 0.4em\relax IEEE, feb 2017, pp. 2--12.

\bibitem{Kikas2017}
R.~Kikas, G.~Gousios, M.~Dumas, and D.~Pfahl, ``{Structure and Evolution of
  Package Dependency Networks},'' in \emph{2017 IEEE/ACM 14th International
  Conference on Mining Software Repositories (MSR)}.\hskip 1em plus 0.5em minus
  0.4em\relax IEEE, may 2017, pp. 102--112.

\bibitem{Constantinou2017}
E.~Constantinou and T.~Mens, ``{Socio-technical evolution of the Ruby ecosystem
  in GitHub},'' in \emph{2017 IEEE 24th International Conference on Software
  Analysis, Evolution and Reengineering (SANER)}.\hskip 1em plus 0.5em minus
  0.4em\relax IEEE, feb 2017, pp. 34--44.

\bibitem{Lum13}
P.~Y. Lum, G.~Singh, A.~Lehman, T.~Ishkanov, M.~Vejdemo-Johansson,
  M.~Alagappan, J.~Carlsson, and G.~Carlsson, ``{Extracting insights from the
  shape of complex data using topology}.''

\bibitem{Costa2017}
J.~P. Costa, ``{The Topological Data Analysis of Time Series Failure Data in
  Software Evolution},'' pp. 25--30, 2017.

\bibitem{SPBG:SPBG07:091-100}
G.~Singh, F.~Memoli, and G.~Carlsson, ``{Topological Methods for the Analysis
  of High Dimensional Data Sets and 3D Object Recognition},'' in
  \emph{Eurographics Symposium on Point-Based Graphics}, M.~Botsch,
  R.~Pajarola, B.~Chen, and M.~Zwicker, Eds.\hskip 1em plus 0.5em minus
  0.4em\relax The Eurographics Association, 2007.

\bibitem{Lungu2008}
M.~Lungu, ``{Towards reverse engineering software ecosystems},'' in \emph{2008
  IEEE International Conference on Software Maintenance}.\hskip 1em plus 0.5em
  minus 0.4em\relax IEEE, sep 2008, pp. 428--431.

\bibitem{Bavota2015}
G.~Bavota, G.~Canfora, M.~{Di Penta}, R.~Oliveto, and S.~Panichella, ``{How the
  Apache community upgrades dependencies: an evolutionary study},''
  \emph{Empirical Software Engineering}, vol.~20, no.~5, pp. 1275--1317, oct
  2015.

\bibitem{DBLP:journals/corr/abs-1301-3781}
T.~Mikolov, K.~Chen, G.~Corrado, and J.~Dean, ``Efficient estimation of word
  representations in vector space,'' \emph{CoRR}, vol. abs/1301.3781, 2013.

\bibitem{Maaten2008}
L.~van~der Maaten and G.~Hinton, ``{Visualizing Data using t-SNE},''
  \emph{Journal of Machine Learning Research}, vol.~9, no. Nov, pp. 2579--2605,
  2008.

\bibitem{Porzio:2008:UAB:1416593.1416598}
G.~C. Porzio, G.~Ragozini, and D.~Vistocco, ``On the use of archetypes as
  benchmarks,'' \emph{Appl. Stoch. Model. Bus. Ind.}, vol.~24, no.~5, pp.
  419--437, Sep. 2008.

\bibitem{Eugster:Leisch:2009:JSSOBK:v30i08}
M.~J.~A. Eugster and F.~Leisch, ``From {Spider-Man} to hero --- archetypal
  analysis in {R},'' \emph{J. of Statistical Softw.}, vol.~30, no.~8, pp.
  1--23, 4 2009.

\bibitem{KulaEMSE2017}
R.~G. Kula, D.~M. German, A.~Ouni, T.~Ishio, and K.~Inoue, ``{Do developers
  update their library dependencies?}'' \emph{Empirical Software Engineering},
  pp. 1--34, may 2017.

\bibitem{Ihara2017}
A.~Ihara, D.~Fujibayashi, H.~Suwa, R.~G. Kula, and K.~Matsumoto,
  ``{Understanding When to Adopt a Library: A Case Study on ASF
  Projects}.''\hskip 1em plus 0.5em minus 0.4em\relax Springer, Cham, may 2017,
  pp. 128--138.

\bibitem{Ishio2017}
T.~Ishio, Y.~Sakaguchi, K.~Ito, and K.~Inoue, ``{Source File Set Search for
  Clone-and-Own Reuse Analysis},'' in \emph{2017 IEEE/ACM 14th International
  Conference on Mining Software Repositories (MSR)}.\hskip 1em plus 0.5em minus
  0.4em\relax IEEE, may 2017, pp. 257--268.

\bibitem{Serebrenik2015}
A.~Serebrenik and T.~Mens, ``{Challenges in Software Ecosystems Research},'' in
  \emph{Proceedings of the 2015 European Conference on Software Architecture
  Workshops - ECSAW '15}.\hskip 1em plus 0.5em minus 0.4em\relax New York, New
  York, USA: ACM Press, 2015, pp. 1--6.

\bibitem{Jansen2013}
S.~Jansen, M.~A. Cusumano, and S.~Brinkkemper, \emph{{Software ecosystems :
  analyzing and managing business networks in the software industry}}.\hskip
  1em plus 0.5em minus 0.4em\relax Edward Elgar, 2013.

\bibitem{German2013}
D.~M. German, B.~Adams, and A.~E. Hassan, ``{The Evolution of the R Software
  Ecosystem},'' in \emph{2013 17th European Conference on Software Maintenance
  and Reengineering}.\hskip 1em plus 0.5em minus 0.4em\relax IEEE, mar 2013,
  pp. 243--252.

\bibitem{Raemaekers2012}
S.~Raemaekers, A.~van Deursen, and J.~Visser, ``{Measuring software library
  stability through historical version analysis},'' in \emph{2012 28th IEEE
  International Conference on Software Maintenance (ICSM)}.\hskip 1em plus
  0.5em minus 0.4em\relax IEEE, sep 2012, pp. 378--387.

\bibitem{KulaVISSOFT2014}
R.~G. Kula, C.~D. Roover, D.~German, T.~Ishio, and K.~Inoue, ``{Visualizing the
  Evolution of Systems and Their Library Dependencies},'' in \emph{2014 Second
  IEEE Working Conference on Software Visualization}.\hskip 1em plus 0.5em
  minus 0.4em\relax IEEE, sep 2014, pp. 127--136.

\bibitem{Yano2015}
Y.~Yano, R.~Kula, T.~Ishio, and K.~Inoue, ``{VerXCombo: An Interactive Data
  Visualization of Popular Library Version Combinations},'' in \emph{IEEE
  International Conference on Program Comprehension}, vol. 2015-Augus, 2015.

\bibitem{Onoue2014}
S.~Onoue, H.~Hata, and k.~Matsumoto, ``{Software Population Pyramids: The
  Current and the Future of OSS Development Communities},'' pp. 34:1--34:4,
  2014.

\bibitem{Onoue2016}
S.~Onoue, H.~Hata, A.~Monden, and K.~Matsumoto, ``{Investigating and projecting
  population structures in open source software projects: A case study of
  projects in GitHub},'' \emph{IEICE Transactions on Information and Systems},
  vol. E99D, no.~5, pp. 1304--1315, 2016.

\bibitem{Nicolau2011}
M.~Nicolau, A.~J. Levine, and G.~Carlsson, ``{Topology based data analysis
  identifies a subgroup of breast cancers with a unique mutational profile and
  excellent survival},'' \emph{Proc. Nat. Acad. Sci.}, p.
  108(17):7265^^e2^^80^^937270, 2011.

\bibitem{Nielson2015}
J.~L. Nielson, J.~Paquette, A.~W. Liu, C.~F. Guandique, C.~A. Tovar, T.~Inoue,
  K.-A. Irvine, J.~C. Gensel, J.~Kloke, T.~C. Petrossian, P.~Y. Lum, G.~E.
  Carlsson, G.~T. Manley, W.~Young, M.~S. Beattie, J.~C. Bresnahan, and A.~R.
  Ferguson, ``{Topological data analysis for discovery in preclinical spinal
  cord injury and traumatic brain injury},'' \emph{Nature Communications},
  vol.~6, 2015.

\bibitem{Louie2016}
A.~Louie, K.~H. Song, A.~Hotson, A.~{Thomas Tate}, D.~S. Schneider, and
  D.~Schneider, ``{How Many Parameters Does It Take to Describe Disease
  Tolerance?}'' \emph{PLOS Biology}, vol.~14, no.~4, p. e1002435, apr 2016.

\bibitem{Ibekwe2014}
A.~M. Ibekwe, J.~Ma, D.~E. Crowley, C.-H. Yang, A.~M. Johnson, T.~C.
  Petrossian, P.~Y. Lum, E.~Franz, and A.~Flieger, ``{Topological data analysis
  of Escherichia coli O157:H7 and non-O157 survival in soils},'' 2014.

\bibitem{Li2015}
L.~Li, W.-Y. Cheng, B.~S. Glicksberg, O.~Gottesman, R.~Tamler, R.~Chen, E.~P.
  Bottinger, and J.~T. Dudley, ``{Identification of type 2 diabetes subgroups
  through topological analysis of patient similarity},'' \emph{Science
  Translational Medicine}, vol.~7, no. 311, pp. 311ra174--311ra174, 2015.

\end{thebibliography}

\end{document}